\documentstyle[12pt]{article}
\def\be{\begin{equation}}
\def\ee{\end{equation}}
\def\bea{\begin{eqnarray}}
\def\eea{\end{eqnarray}}

\begin{document}

\begin{center}
{\Large{\bf Mixed Branes Interaction in Compact Spacetime }}
\vskip .5cm
{\large H. Arfaei and D. Kamani}
\vskip .1cm
 {\it Institute for Studies in Theoretical Physics and 
Mathematics
\\ Tehran P.O.Box: 19395-5531, Iran}\\
{\it and}\\
{\it Department of physics, Sharif University of Technology
P.O.Box: 11365-9161}\\
{\sl e-mail:arfaei@theory.ipm.ac.ir\\
e-mail:kamani@theory.ipm.ac.ir}
\\
\end{center}

\begin{abstract} 
We present a general description of two mixed branes interactions. For  
this we consider two mixed branes with dimensions $p_1$ and $p_2$,
 in external field $B_{\mu\nu}$ and arbitrary gauge fields 
 $A^{(1)}_{\alpha_1}$ and $A^{(2)}_{\alpha_2}$ on the world volume of them,   
 in spacetime in which some of its directions are compactified on  
 circles with different radii. Some examples are considered to clear these   
 general interactions. Finally contribution of the massless states on the 
 interactions is extracted. Closed string with
 mixed boundary conditions and boundary state formalism, provide useful  
 tools for calculation of these interactions. 
\end{abstract} 
\newpage
\section{Introduction}

A way of describing D-branes is boundary state formalism \cite{1,2,
3,4,5,6,7}. The boundary state can be interpreted as a source for a closed
string emitted by a $D$-brane. Thus the interaction of two $D$-branes
is viewed as an exchange of closed string states, therefore it is 
computed with a tree level diagram in which two boundary states are connected
by means of a closed string propagator.

By introducing back-ground fields $B_{\mu \nu}$ and $A_{\alpha}$ a 
$U(1)$ gauge field (which lives in the $D$-brane) in the string
$\sigma$-model one obtains mixed boundary conditions for string, these
fields appear in the boundary states and modify the tensions 
of the branes. Mixed boundary state    
formalism enables us to consider the problems not easily accessible    
to the canonical approach via open strings.

Mixed boundary conditions have been used for studying properties of
$D$-branes in back-ground fields \cite{8,9,10,11,12,13,14,15,16}. 
In Ref.\cite{15},the interaction between $D_0$ and $D_6$ branes with
back-ground gauge fields has been discussed.
In Ref.\cite{16} we applied mixed boundary conditions for the closed bosonic 
string and studied the interactions of the branes in spacetime with 
compactification on tori. Inclusion of fermionic degrees of freedom is
non trivial and requires its own techniques. This is what we take on
in this article.

We use the covariant formalism to extract the boundary states
which now involve apart from the bosonic and fermionic components, due to the 
covariance the ghosts and superghosts elements. Then we
compute the interaction amplitude between two mixed branes 
with arbitrary dimensions $p_1$ and $p_2$ and field strengths ${\cal{F}}_1$ 
and ${\cal{F}}_2$ as a closed string tree 
level diagram. Then we proceed to study the above considerations 
when certain directions are compactified. Finally to elucidate our
general computations we apply our results to special cases. It is worth
emphasizing that part of these special cases are either inaccessible
to the canonical methods and the other part are very difficult to 
handle by canonical formulation. 
Among the special cases that
will be considered is the parallel $m_{p_1}$ and  
$m_{p_2}$-branes with the same total field strength.
The $NS \otimes NS$ sector interaction for $p_2-p_1=4$ vanishes, 
also for $p_1=p_2$ the total
interaction vanishes. Other examples include different internal fields 
are : parallel $m_1-m_{1'}$ 
, perpendicular $m_1-m_{1'}$, $m_2-m_0$, parallel $m_2-m_{2'}$, 
perpendicular $m_2-m_{2'}$ and parallel $m_5-m_1$ branes. They are considered 
to clear more properties of the field strengths and compactification 
effects on interaction amplitude. Finally contribution of the massless states
on the amplitude for the NS-NS and R-R sectors separately will be 
obtained. In this article we denote a mixed brane with 
dimension ``$p$'' by notation ``$m_p$-brane''. 


\section{Boundary state}
First we develop the boundary state formalism for the branes with background
gauge fields. Deriving the boundary conditions from a $\sigma$-model action,
we turn them in to boundary state equations which we will solve in the next 
subsection.

\subsection{Boundary state equations}

A ${\sigma}$-model action with $B_{\mu\nu}$ field and two boundary terms 
\cite{17} corresponding to the two $m_{p_1}$ and $m_{p_2}$-branes gauge  
fields is
\bea
S = -\frac{1}{4\pi \alpha'} \int_{\Sigma} d^2 \sigma \bigg{(} 
\sqrt{-g} g^{ab}G_{\mu \nu}\partial_a X^{\mu}\partial_b X^{\nu}+
\epsilon^{ab}B_{\mu \nu}\partial_a X^{\mu}\partial_b X^{\nu}\bigg{)}
\nonumber\\
-\frac{1}{2\pi \alpha'}\int_{(\partial \Sigma)_1}d\sigma
A^{(1)}_{\alpha_1}\partial_{\sigma} X^{\alpha_1} + \frac{1}{2\pi \alpha'}
\int_{(\partial \Sigma)_2}d\sigma A^{(2)}_{\alpha_2}\partial_{\sigma}
X^{\alpha_2}\;\;,
\eea
where $\Sigma$ is the world sheet of closed string exchanged between the 
branes and $(\partial \Sigma)
_1$ and $(\partial \Sigma)_2$ are two boundaries of the world sheet.  
The first boundary is at $\tau = 0$ and the second at $\tau = \tau_0$. 
The two $U(1)$ gauge fields $A^{(1)}_{\alpha_1}$ and $A^{(2)}_{\alpha_2}$ 
live in $m_{p_1}$ and $m_{p_2}$-branes respectively. 
$\alpha_1 , \beta_1 \in \{0,
{\bar \alpha}_1,...,{\bar \alpha}_{p_1}\}$, this set shows directions 
along the $m_{p_1}$-brane and $\{i_1\}$ show directions perpendicular to 
it. Likely $\alpha_2, \beta_2 \in \{0,{\bar \beta}_1,  
...,{\bar \beta}_{p_2} \}$ and $\{i_2\}$ for $m_{p_2}$-brane. 
$G_{\mu \nu}$ and $B_{\mu \nu}$ are usual back-ground fields. Let 
$B_{\mu \nu}(X)$ and $G_{\mu \nu}(X)$ be constant fields. Variation 
of this action with respect to $X^{\mu}(\sigma, \tau)$ gives the boundary 
state equations and equation of motion of $X^{\mu}(\sigma , \tau)$.
Using the convention $\epsilon^{01} = -\epsilon^{10} = 1$, we obtain
\bea
&~&\bigg{(}\partial_{\tau} X^{\alpha_1}+{\cal{F}}
^{\alpha_1}_{(1)\;\; \beta_1}
\partial_{\sigma}X^{\beta_1}- B^{\alpha_1}\;_{j_1}\partial_{\sigma}
X^{j_1}\bigg{)}_{\tau=0}\mid B^1_x \rangle = 0\;\;,
\nonumber\\
&~&(\delta X^{i_1})_{\tau = 0} \mid B^1_x \rangle = 0\;\;,
\nonumber\\
&~&\bigg{(}\partial_{\tau} X^{\alpha_2}+{\cal{F}}
^{\alpha_2}_{(2)\;\; \beta_2}
\partial_{\sigma} X^{\beta_2}-B^{\alpha_2}\;_{j_2}\partial
_{\sigma}X^{j_2}\bigg{)}_{\tau_0} \mid B^2_x \rangle = 0\;\;,
\nonumber\\
&~&(\delta X^{i_2})_{\tau_0} \mid B^2_x \rangle = 0\;\;,
\eea
where
\bea
&~&{\cal{F}}_{(1) \alpha_1 \beta_1} \equiv \partial_{\alpha_1} 
A^{(1)}_{\beta_1 }- \partial_{\beta_1}A^{(1)}_{\alpha_1}
-B_{\alpha_1 \beta_1}\;\;,
\nonumber\\
&~&{\cal{F}}_{(2) \alpha_2 \beta_2} \equiv \partial_{\alpha_2}
A^{(2)}_{\beta_2 }- \partial_{\beta_2}A^{(2)}_{\alpha_2}
-B_{\alpha_2 \beta_2}\;\;.
\eea
The transverse coordinates of the two branes 
$\{ y_1 ^{i_1} \}$ and $\{ y_2 ^{i_2} \}$ are kept fixed $i.e$
\bea
&~&[X^{i_1}(\sigma , \tau )-y^{i_1}_1]_{\tau =0}\mid B^1_x \rangle = 0\;\;,
\nonumber\\
&~&[X^{i_2}(\sigma , \tau )-y^{i_2}_2]_{\tau_0} \mid B^2_x \rangle = 0\;\;. 
\eea
These imply $\partial_{\sigma}X^{j_1}$ (, 
$\partial_{\sigma}X^{j_2}$ ) vanish and be dropped from the first
(third) equation of (2).

Solution of the equations of motion of the closed string is
\bea
X^{\mu}(\sigma , \tau) = x^{\mu}+2\alpha' p^{\mu}\tau+2L^{\mu}\sigma 
+\frac{i}{2}\sqrt{2\alpha'}\sum_{m\neq 0}\frac{1}{m}\bigg{(}
\alpha^{\mu}_m e^{-2im(\tau-\sigma)}+{\tilde \alpha}^{\mu}_m 
e^{-2im(\tau+\sigma)}\bigg{)}\;\;,
\eea
where $L^{\mu}$ is zero for non compact directions. For compact 
directions we have $L^{\mu}=N^\mu R^\mu$ and $p^\mu=\frac{M^\mu}{R^\mu}$, 
in which $N^\mu$ is the winding number and $M^\mu$ is the momentum number 
of closed string state, and $R_{\mu}$ is the radius of compactification of  
$X^{\mu}$-direction. Combining the solution of the equation of motion 
and the boundary state equations we obtain 
\bea
\bigg{(}p^{\alpha_2}  + \frac{1}{\alpha'}{\cal{F}}^{\alpha_2}
_{(2)\;\; \beta_2}L^{\beta_2} \bigg{)} \mid B^2_x , \tau_0 \rangle =0\;\;,
\eea
\bea
\bigg{(} (1-{\cal{F}}_2)^{\alpha_2}\;_{\beta_2}\;
\alpha^{\beta_2}_n e^{-2in\tau_0}+(1+{\cal{F}}_2)^{\alpha_2}
\;_{\beta_2}\;{\tilde \alpha}^{\beta_2}_{-n} e^{2in\tau_0} \bigg{)}
\mid B^2_x , \tau_0 \rangle = 0\;\;,
\eea
\bea
\bigg{(}\alpha^{i_2}_n e^{-2in\tau_0} -{\tilde \alpha}^{i_2}_{-n}
e^{2in\tau_0} \bigg{)} \mid B^2_x , \tau_0 \rangle = 0\;\;,
\eea
\bea
(x^{i_2}+2\alpha' p^{i_2} \tau_0  -y^{i_2}_2)
\mid B^2_x , \tau_0 \rangle =0\;\;,
\eea
\bea
L^{i_2} \mid B^2_x , \tau_0 \rangle = 0\;\;,
\eea
The boundary conditions on the fermionic degrees of freedom should be 
imposed on both $R \otimes R$ and $NS \otimes NS$ sectors.
World sheet supersymmetry requires the two sectors to 
satisfy the boundary conditions,                   
\bea
\bigg{[}(\psi^{\alpha_2}-i\eta_2 {\tilde \psi}^{\alpha_2})- 
{\cal{F}}^{\alpha_2}_{(2)\;\;\beta_2}\;(\psi^{\beta_2}+i\eta_2 
{\tilde \psi}^{\beta_2}) \bigg{]}_{\tau_0} \mid B^2_{\psi} , 
\eta_2 , \tau_0 \rangle =0\;\;,
\eea
\bea
(\psi^{i_2}+i\eta_2 {\tilde \psi}^{i_2} )_{\tau_0} \mid B^2_{\psi} 
, \eta_2 , \tau_0 \rangle = 0\;\;, 
\eea
where $\eta_2 = \pm 1$ is the phase used to make GSO projection easily. 
 These states preserve half of the world sheet supersymmetry.
Expanding the fermions in Fourier modes the boundary conditions
become
\bea
\bigg{(}\psi ^{i_2} _k +i\eta_2 e^{4ik\tau_0} {\tilde \psi}^{i_2}_{-k}
\bigg{)} \mid B^2_{\psi} , \eta_2 , \tau_0 \rangle = 0\;\;,
\eea
\bea
\bigg{(}\psi ^{\alpha_2} _k - i\eta_2 Q^{\alpha_2}_{(2)\;\; \beta_2}
\;e^{4ik\tau_0} {\tilde \psi}^{\beta_2}_{-k} \bigg{)} \mid B^2_{\psi} , 
\eta_2 , \tau_0 \rangle = 0\;\;,
\eea
where the index $k$ is integer in the R-R sector and half-integer 
in the NS-NS sector. The matrix $Q_2$ is 
\bea
Q_2 \equiv (1-{\cal{F}}_2)^{-1}(1+{\cal{F}}_2)\;\;,
\eea
since ${\cal{F}}_2$ is antisymmetric , $Q_2$ is an orthogonal matrix. 
Since we use the covariant formalism we shall introduce ghost for the  
bosonic and fermionic gauge (reparametrization ) degrees of freedom. Let the 
ghost coordinates be $b(\sigma , \tau)$ 
and $c(\sigma , \tau)$. Vanishing of the variation of the ghosts action 
gives the ghosts boundary conditions. Therefore 
the ghost modes satisfy the following boundary conditions
\bea
\bigg{(}b_n\; e^{-2in\tau_0} - {\tilde b}_{-n}\; e^{2in\tau_0} \bigg{)}
\mid B^2_{gh} , \tau_0 \rangle = 0\;\;,
\eea
\bea
\bigg{(}c_n\; e^{-2in\tau_0}+{\tilde c}_{-n}\; e^{2in\tau_0} \bigg{)} 
\mid B^2_{gh} , \tau_0 \rangle = 0 \;\;,
\eea
where $n$ is non zero integer. 
The same consideration also determine the boundary conditions for superghost
coordinates $\beta$,$\tilde \beta$,$\gamma $, ${\tilde \gamma}$ 
\bea
\bigg{(}\gamma_k + i\eta_2 {\tilde \gamma}_{-k}\; e^{4ik\tau_0}\bigg{)}
\mid B^2_{sgh} , \eta_2 , \tau_0 \rangle = 0 \;\;,
\eea
\bea
\bigg{(}\beta_k + i\eta_2 {\tilde \beta}_{-k}\;e^{4ik\tau_0} \bigg{)} 
\mid B^2_{sgh} , \eta_2 , \tau_0 \rangle = 0\;\;.
\eea
The index $k$ as previous is integer in the R-R sector and half-integer 
in the NS-NS sector.
\subsection{Solutions of boundary state equations} 

To find the boundary states we shall proceed to solve the equations 6-10,
13-14 and 16-19.
Equations (6-10) have the solution,  
\bea
\mid B^2_x , \tau_0 \rangle = \sum_{\{p^{\alpha_2}\}} \mid B_x^2, \tau_0
, p^0 , p^{{\bar \beta}_1} , ... , p^{{\bar \beta}_{p_2}} \rangle \;\;,
\eea
\bea
\mid B^2_x , \tau_0 , p^0 , p^{{\bar \beta}_1} , ... ,
p^{{\bar \beta}_{p_2}}
\rangle = \frac{T_{p_2}}{2}\sqrt{\det(1-{\cal{F}}_2)}\;e^{i\alpha' \tau_0
\sum_{i_2}(p^{i_2}_{op})^2}\delta^{(d-p_2-1)}(x^{i_2}-y^{i_2}_2)
\nonumber\\
\times \exp \bigg{[} {-\sum_{m=1}^{\infty}\frac{1}{m}e^{4im\tau_0} 
\alpha^{\mu}_{-m}
S^{(2)}_{\mu \nu}{\tilde \alpha}^{\nu}_{-m}}\bigg{]}
\mid 0 \rangle \prod_{i_2} \mid p^{i_2}_L = p^{i_2}_R = 0 \rangle 
\prod_{\alpha_2} \mid p^{\alpha_2} \rangle \;\;,
\eea
where the matrix $S^{\mu}_{(2)\; \nu}$ is 
\bea
S^{\mu}_{(2)\; \nu} = (Q^{\alpha_2}_{(2)\; \beta_2} \; , \; -\delta^{i_2}
\;_{j_2}) \;\;,
\eea
and $T_{p_2}$ is a constant depending on the tension of $m_{p_2}$-brane
\cite{3,6}. The overall factor $\sqrt{\det(1-{\cal{F}}_2)}$ is 
expected by the path integral with boundary action \cite{13,18}. 
In (20) the summation 
over $\{p^{\alpha_2}\}$ can change to a sum over winding numbers 
$\{N^{\alpha_{2c}}\}$ due to the equation (6) which implies
\bea
p^{\alpha_2}=-\frac{1}{\alpha'} \sum_{\beta_{2c}}{\cal{F}}^{\alpha_2}
_{(2)\;\; \beta_{2c}}\;{\ell}^{\beta_{2c}} \;\;, 
\eea
 where ${\ell}^{\beta_{2c}} = 
N^{\beta_{2c}}R^{\beta_{2c}}$ and $\beta_{2c}$ shows the direction along 
$m_{p_2}$-brane which is compact. This relation implies that the 
closed string state   
can have non zero momentum along the world brane if there are non zero
back-ground internal gauge fields and at least one of the brane
directions is compact. This relation correlates the
momentum of closed string state along the brane directions to its 
winding numbers. For compact directions of the brane, the closed 
string state also has momentum
numbers $\{M^{\alpha_{2c}}\}$, therefore when $\alpha_2$ in (23) refers
to the compact directions of brane, we have
\bea
\frac{M^{\alpha_{2c}}}{R^{\alpha_{2c}}}=-\frac{1}{\alpha'}
\sum_{\beta_{2c}}{\cal{F}}^{\alpha_{2c}}_{(2) \;\beta_{2c}}
R^{\beta_{2c}}N^{\beta_{2c}} \;\;,
\eea
this is a relation between momentum numbers and winding numbers of a 
given closed string state, more details can be found in \cite{16} where
the pure bosonic case is discussed.

Ghost part of boundary state has the form 
\bea
\mid B^2_{gh} , \tau_0 \rangle = \exp \bigg{[}{\sum_{m=1}^{\infty}
e^{4im\tau_0}(c_{-m}{\tilde b}_{-m}-b_{-m}{\tilde c}_{-m})}\bigg{]}
\frac{c_0+\tilde{c}_0}{2}\mid q=1 \rangle \mid \tilde{q}=1 \rangle \;\;, 
\eea
Let us denote the fermionic modes in the R-R sector with $d_n ^\mu$ and
in the NS-NS sector with $b_r ^\mu$, therefore
the fermionic and the superghost parts of the NS-NS sector boundary 
state in the $(-1 , -1)$ picture is
\bea
\mid B^2_{\psi} , \eta_2 , \tau_0 \rangle _{NS} = \exp \bigg{[}
{i\eta_2 \sum_{r=1/2}
^{\infty}e^{4ir\tau_0}b^{\mu}_{-r}S^{(2)}_{\mu \nu}{\tilde b}^{\nu}
_{-r}} \bigg{]} \mid 0 \rangle \;\;,
\eea
\bea
\mid B^2_{sgh} , \eta_2 , \tau_0 \rangle _{NS} = \exp \bigg{[}
{i\eta_2 \sum_{r=1/2}
^{\infty}e^{4ir\tau_0}(\gamma_{-r}{\tilde \beta}_{-r} - \beta_{-r}
{\tilde \gamma}_{-r})} \bigg{]}\mid P=-1 , \tilde P = -1 \rangle \;\;.
\eea
The fermionic and the superghost parts of the R-R sector boundary 
state in the $(-1/2 , -3/2)$ picture is 
\bea
\mid B^2_{\psi} , \eta_2 , \tau_0 \rangle _R= 
\frac{1}{\sqrt{\det(1-{\cal{F}}_2)}}\;\exp \bigg{[}{i\eta_2 \sum_{m=1}
^{\infty}e^{4im\tau_0}d^{\mu}_{-m}S^{(2)}_{\mu \nu}{\tilde d}^{\nu}
_{-m}}\bigg{]} \mid B^2_{\psi} , \eta_2  \rangle ^{(0)}_R \;\;,
\eea
\bea
\mid B^2_{sgh} , \eta_2 , \tau_0 \rangle _R = \exp \bigg{[}
{i\eta_2 \sum_{m=1}
^{\infty}e^{4im\tau_0}(\gamma_{-m}{\tilde \beta}_{-m} - \beta_{-m}
{\tilde \gamma}_{-m})+i\eta_2 \gamma _0 {\tilde \beta}_0}\bigg{]} 
\mid P=-1/2 , {\tilde P} = -3/2 \rangle \;\;, 
\eea
where the superghost vacuum is in the $(-1/2 , -3/2)$ 
picture and is annihilated 
by $\beta_0$ and ${\tilde \gamma}_0$ \cite{19} and $\mid B^2_{\psi} , 
\eta_2 \rangle ^{(0)}_R $ is the fermionic zero mode boundary state.
Appearance of the determinant in the denominator is the consequence of the 
path integral over the fermions with fermionic boundary term.
Comparison of (21) and (28) implies that in the R-R sector the 
normalizing determinant factors of the 
bosonic boundary determinant and its fermionic partner cancel. However
this factor remains in the NS-NS sector.

We now derive the explicit form of $\mid B^2_{\psi} , \eta_2 \rangle
^{(0)}_R$ both in type IIA and type IIB theories.  
It obeys the equations (13) and (14) with $k = 0$, $i.e.$
\bea
(d^{i_2}_0 + i\eta_2 {\tilde d}^{i_2}_0) \mid B^2_{\psi} , \eta_2
\rangle ^{(0)}_R = 0 \;\;,
\eea
\bea
(d^{\alpha_2}_0 - i\eta_2 Q^{\alpha_2}_{(2)\;\;\beta_2}\;{\tilde d}
^{\beta_2}_0 ) \mid B^2_{\psi} , \eta_2 \rangle ^{(0)}_R = 0 \;\;,
\eea
or in combined form,
\bea
(d^{\mu}_0 - i\eta_2 S^{\mu}_{(2)\;\;\nu}{\tilde d}^{\nu}_0
) \mid B^2_{\psi} , \eta_2 \rangle ^{(0)}_R = 0 \;\;.
\eea
The vacuum for the fermionic zero modes $d^{\mu}_0$ and ${\tilde d}
^{\mu}_0$ can be written as \cite{6} 
\bea
\mid A \rangle \mid {\tilde B} \rangle = 
\lim_{z,{\bar z} \rightarrow 0} 
S^A (z) {\tilde S}^B (\bar z) \mid 0 \rangle \;\;, 
\eea
where $S^A$ and ${\tilde S}^B $ are the spin fields in the 32-dimensional
Majorana representation. We use a chiral representation for the 
$32 \times 32 \;\Gamma$-matrices of SO(1,9) as in reference \cite{6}.  
Also the action of the Ramond oscillators
$d^{\mu}_n$ and ${\tilde d}^{\mu}_n$ on the state $\mid A \rangle
\mid {\tilde B} \rangle $ are given in \cite{6}, therefore we consider 
solution of (32) of the form (like \cite{7}), 
\bea
\mid B^2_{\psi} , \eta_2 \rangle ^{(0)}_R = {\cal{M}}^{(\eta_2)}_{AB}
\mid A \rangle \mid {\tilde B} \rangle \;\;,
\eea
therefore the $32 \times 32$ matrix ${\cal{M}}^{(\eta_2)}$ satisfies
the following equation 
\bea
(\Gamma^{\mu})^T {\cal{M}}^{(\eta_2)}-i\eta_2 S^{\mu}_{(2)\;\;\nu}
\Gamma_{11} {\cal{M}}^{(\eta_2)}\Gamma^{\nu} = 0 \;\;.
\eea
For this equation we consider a solution with the form 
\bea
{\cal{M}}^{(\eta_2)} = C\Gamma^0 \Gamma^{{\bar \beta}_1} ... \Gamma^{  
{\bar \beta}_{p_2}}\bigg{(}\frac{1+i\eta_2 \Gamma_{11}}
{1+i\eta_2}\bigg{)}G_2 \;\;,
\eea
where $C$ is the charge conjugation matrix and ${\bar \beta}_i$'s show
the space directions of the $m_{p_2}$-brane world volume. For the case  
of ${\cal{F}}_2 = 0$, $G_2$ must be equal to the unit 
matrix, in this case (36) reduces to the equation (2.22) of \cite{7}. 
From (35) and (36) we see that $G_2$ must satisfy the equation 
\bea
\Gamma^{\alpha}G_2 = Q^{\alpha}_{(2)\;\; \beta}\;G_2 \Gamma^{\beta}
\;\;\;\;\; ,\;\;\;\;\; \alpha , \beta \in \{0, {\bar \beta}_1 , ... ,
{\bar \beta}_{p_2} \} \;\;.
\eea
Therefore matrix $G_2$ has the solution with the conventional form
\bea
G_2 = e^{\frac{1}{2}({\cal{F}}_2)
_{\alpha \beta}\Gamma^{\alpha}\Gamma^{\beta}} \;\;, 
\eea
Indeed one must expand the exponential with the convention that all
gamma matrices anti commute, therefore there are a
finite number of terms. This convention is in Ref.\cite{11,13}. 
For example for $m_p$-brane with 
$p=1$ along $X^1$, $p=2$ along
$(X^1 , X^2)$ and $p=3$ along $(X^1 , X^2 , X^3)$ directions, respectively 
we have
\bea
G_2 = 1+{\cal{F}}_{(2) 01} \Gamma^0 \Gamma^1 \;\;, 
\eea
\bea
G_2 = 1+{\cal{F}}_{(2) 01} \Gamma^0 \Gamma^1 + {\cal{F}}_{(2) 02}  
\Gamma^0 \Gamma^2 + {\cal{F}}_{(2) 12}\Gamma^1 \Gamma^2 \;\;,
\eea
\bea
G_2 = 1+\frac{1}{2} {\cal{F}}_{(2) \alpha \beta} \Gamma^{\alpha}
\Gamma^\beta + ({\cal{F}}_{(2) 01}{\cal{F}}_{(2) 23}- {\cal{F}}_{(2) 02} 
{\cal{F}}_{(2) 13}+{\cal{F}}_{(2) 03} {\cal{F}}_{(2) 12}) \Gamma^0
\Gamma^1 \Gamma^2 \Gamma^3 \;\;,  
\eea
and $\alpha , \beta = 0,1,2,3$.

This special representation of $\Gamma$-matrices allow us to decompose
the spinors in chiral and anti-chiral components ($A = (\alpha ,  
{\dot \alpha})$) with sixteen dimensional indices $\alpha$ and ${\dot 
\alpha}$. In the type IIA theory $p_2$ is even, therefore ${\cal{M}}^
{(\eta_2)}$ is a block-diagonal matrix, whereas in the type IIB theory
$p_2$ is odd and therefore ${\cal{M}}^{(\eta_2)}$ is in the form of 
an off diagonal matrix with matrices as its elements. 
Thus in the sixteen-dimensional notation, $\mid 
B^2_{\psi} , \eta_2 \rangle ^{(0)}_R$ becomes 
\bea
\mid B^2_{\psi} , \eta_2 \rangle ^{(0)}_R = M_{\alpha \beta}
\mid \alpha \rangle _{-1/2} \mid {\tilde \beta} \rangle _{-3/2}
-i\eta_2 M_{{\dot \alpha}{\dot \beta}} \mid {\dot \alpha} \rangle
_{-1/2} \mid {\tilde {\dot \beta}} \rangle _{-3/2} \;\;\;\;\;for IIA \;\;,
\eea
\bea
\mid B^2_{\psi} , \eta_2 \rangle ^{(0)}_R = M_{{\dot \alpha} \beta}
\mid {\dot \alpha} \rangle _{-1/2} \mid {\tilde \beta} \rangle _{-3/2}
-i\eta_2 M_{\alpha {\dot \beta}} \mid \alpha \rangle _{-1/2} 
\mid {\tilde{\dot \beta}} \rangle _{-3/2} \;\;\;\;\;\;for IIB \;\;,
\eea
where the matrix $M_{AB}$ has definition
\bea
{M_{AB}} \equiv \left( \begin{array}{cc}
M_{\alpha \beta}&M_{\alpha {\dot \beta}} \\
M_{{\dot \alpha} \beta}&M_{{\dot \alpha}{\dot \beta}}
\end{array} \right)
= ( C \Gamma^0 \Gamma^{{\bar \beta}_1} ... \Gamma^{{\bar \beta}
 _{p_2}}G_2 ) _{AB} \;\;.
\eea
\subsection{GSO projection of the boundary state}
For both NS-NS and R-R sectors the complete boundary state can be written 
as the following product 
\bea
\mid B_2 , \eta_2 , \tau_0 \rangle _{R,NS} = \;\mid B^2_x , \tau_0 \rangle 
\mid B^2_{gh} , \tau_0 \rangle \mid B^2_{\psi} , \eta_2 , \tau_0
\rangle _{R,NS} \mid B^2_{sgh} , \eta_2 , \tau_0 \rangle _{R,NS} \;\;.
\eea
The projected boundary state in the NS-NS sector is \cite{7}
\bea
\mid B_2 , \tau_0 \rangle _{NS} = \frac{1-(-1)^{F+G}}{2}
\frac{1-(-1)^{{\tilde F}+{\tilde G}}}{2} \mid B_2 , + , \tau_0 
\rangle _{NS} \;\;,
\eea
where $F$ and $G$ are 
\bea
F = \sum_{r=1/2}^{\infty} b^{\mu}_{-r} b_{r \mu}\;\;\;\;\;\;
, \;\;\;\; G = -\sum_{r=1/2}^{\infty}(\gamma_{-r} \beta_r + 
\beta_{-r} \gamma_r)\;\;.
\eea
Similar definitions hold for ${\tilde F}$ and ${\tilde G}$. 
Therefore projected state becomes
\bea
\mid B_2 , \tau_0 \rangle _{NS} = \frac{1}{2} ( \mid B_2 , + , \tau_0
 \rangle _{NS} - \mid B_2 , - , \tau_0 \rangle _{NS}\; ) \;\;.
\eea
 In the R-R sector the projection is
\bea
\mid B_2 , \tau_0 \rangle _R = \frac{1 + (-1)^p (-1)^{F+G}}{2}
\frac{1 - (-1)^{{\tilde F}+{\tilde G}}}{2} \mid B_2 , + , \tau_0
\rangle _R \;\;,
\eea
where $p$ is even for type IIA and odd for type IIB, and  
\bea
(-1)^F = \Gamma_{11} (-1)^{\sum_{m=1}^{\infty}d^{\mu}_{-m}d_{m \mu}} 
\;\;\;\; ,\;\;\;\; G = -\gamma_0 \beta_0 -\sum_{m=1}^{\infty}
(\gamma_{-m} \beta_m + \beta_{-m} \gamma_m ) \;\;.
\eea
Finally the projected state is
\bea
\mid B_2 , \tau_0 \rangle _R = \frac{1}{2} (\mid B_2 , + , \tau_0 \rangle
_R + \mid B_2 , - , \tau_0 \rangle _R\; ) \;\;.
\eea
Equation (48) and (51) are similar to the case in which ${\cal{F}}_2 = 0$.
\section{Mixed branes interaction}
Before calculation of the interaction amplitude, let us  
introduce some notations
for the positions of these two mixed branes. The set $\{i\}$ shows 
indices for directions perpendicular to the both of the branes, $\{u\}$   
for the directions along the both of them, $\{\alpha'_1\}$ for directions  
along $m_{p_1}$ and perpendicular to the $m_{p_2}$ and $\{\alpha'_2\}$
for directions along $m_{p_2}$ and perpendicular to the $m_{p_1}$ 
-branes. It can be seen that for example $\{i_1\} = \{i\} \bigcup 
\{\alpha'_2 \}$, $\{ \alpha_1 \}=\{u\} \bigcup \{ \alpha'_1 \}$.
\subsection{The amplitude for the NS-NS sector}

For calculation of the amplitude we need to the conjugate form of the  
boundary states. In the NS-NS sector there are 
\bea
_{NS} \langle B^1_{\psi} , \eta_1 \mid \;= \langle 0 \mid e^{-i\eta_1 
\sum_{r=1/2}^{\infty}b^{\mu}_rS^{(1)}_{\mu \nu}{\tilde b}^{\nu}_r} \;\;, 
\eea
\bea
_{NS} \langle B^1_{sgh} , \eta_1 \mid \;= \langle P=-1 , {\tilde P}
=-1 \mid e^{-i\eta_1 \sum_{r=1/2}^{\infty} ({\tilde \beta}_r 
\gamma_r - {\tilde \gamma}_r \beta_r)}\;\;.
\eea
The two mixed branes simply interact via exchange of closed strings, 
and the amplitude is
\bea
{\cal{A}} = \langle B_1 \mid D \mid B_2 , \tau_0 = 0 \rangle \;\;,
\eea
where ``$D$'' is closed string propagator and one must use 
the GSO projected boundary states.
The NS-NS sector amplitude becomes
\bea
{\cal{A}}_{NS-NS} &=& \frac{T_{p_1}T_{p_2}}{8(2\pi)^{d_i}}\alpha'
\sqrt{\det(1-{\cal{F}}_1)\det(1-{\cal{F}}_2)}
\int_0^\infty dt \bigg{\{} \bigg{(} 
\sqrt{\frac{\pi}{\alpha't}} \;\bigg{)}^{d_{i_n}} 
\nonumber\\
&~& \times e^{-\frac{1}{4\alpha't}\sum_{i_n}(y^{i_n}_1
-y^{i_n}_2)^2} \prod_{i_c}\Theta_3 \bigg{(} \frac{y^{i_c}_1-y^{i_c}_2}
{2\pi R_{i_c}} \mid \frac{i\alpha't}{\pi (R_{i_c})^2}
\bigg{)}
\nonumber\\
&~&\times\frac{1}{q} \bigg{(} \prod_{n=1}^{\infty} \bigg{[} \bigg{(}
\frac{1-q^{2n}}{1+q^{2n-1}} \bigg{)}^2 \;\frac{\det(1+S_1S^T_2 q^{2n-1})}
{\det(1-S_1S^T_2q^{2n})} \bigg{]}
\nonumber\\
&~&-\prod_{n=1}^{\infty} \bigg{[} \bigg{(} \frac{1-q^{2n}}{1-q^{2n-1}}
\bigg{)}^2 \;\frac{\det(1-S_1S^T_2q^{2n-1})}{\det(1-S_1S^T_2q^{2n})}
\bigg{]} \bigg{)}
\nonumber\\
&~&\times \sum_{\{N^{u_c}\}} \bigg{[} (2\pi)^{d_u} \prod_u [\delta(p^u_1
-p^u_2)]\; \exp [\frac{i}{\alpha'}{\ell}^{u_c}({\cal{F}}^{\alpha'_1}
_{(1)\;\; u_c}
y^{\alpha'_1}_2 -{\cal{F}}^{\alpha'_2}_{(2)\;\;u_c}y^{\alpha'_2}_1)]
\nonumber\\
&~&\times \exp[-\frac{t}{\alpha'}{\ell}^{u_c}{\ell}^{v_c}
( \eta_{u_c v_c}+{\cal{F}}^u _{(1)\;u_c} {\cal{F}}_{(2)\;uv_c} 
+{\cal{F}}^{\alpha'_1}_{(1)\;u_c}{\cal{F}}^{\alpha'_1}_{(1)\;v_c}
+{\cal{F}}^{\alpha'_2}_{(2)\;u_c}{\cal{F}}^{\alpha'_2}_{(2)
\;v_c}) ] \bigg{]} \bigg{\}}  
\eea
where $q = e^{-2t}$. In this formula $p^u_1 = -\frac{1}{\alpha'} 
{\cal{F}}^u_{(1)\;\;v_c}N^{v_c}R^{v_c}$ and $p^u_2 = -\frac{1}{\alpha'}
{\cal{F}}^u_{(2)\;\;v_c}N^{v_c}R^{v_c}$. 
Indices $\{u_c , v_c , ... \}$ show
compact part of $\{u\}$ , $d_u$ and $d_i$ are dimensions of 
$\{X^u\}$ and $\{X^i\}$ respectively. Also
$\{i_n\}$ and $\{i_c\}$ are non compact part and compact
part of $\{i\}$ region respectively. ${\ell}^{u_c}$ as 
previous is $N^{u_c}R^{u_c}$.
 Note that determinant in the denominators comes from the world sheet  
bosons and in the numerators from the fermions. This amplitude is symmetric 
under the exchange of the indices ``1'' and ``2'', $i.e$ ${\cal{A}}_{NS}
(1,2)={\cal{A}}_{NS}^*(2,1)$ as expected.
In this amplitude
we see how the effects of compactification appear. Later we will 
see that this compactification structure will be repeated in the
R-R sector.

The momentum delta functions put severe restrictions on the summation.
The term corresponding to $N^{u_c} = 0 $
for all $u_c$, gives $p_1^u = p_2^u = 0$ and
is always present. Other terms occur only if the
two internal back-ground fields and radii of compactification 
with some sets $\{N^{u_c}\}$ satisfy
the relation $\sum_{v_c}({\cal{F}}^u_{(1)\;\;v_c}N^{v_c}R^{v_c}) 
=\sum_{v_c}({\cal{F}}^u
_{(2)\;\;v_c}N^{v_c}R^{v_c}) $ for all $u$.

Now suppose there is no compact direction, then (55) simplifies,
\bea
{\cal{A}}^{(nc)}_{NS-NS} &=& \frac{T_{p_1}T_{p_2}}{8(2\pi)^{d_i}}
\alpha' V_u \sqrt{\det(1-{\cal{F}}_1)\det(1-{\cal{F}}_2)} \int_0^\infty
dt \bigg{\{} \bigg{(} \sqrt{\frac{\pi}{\alpha't}}\;\bigg{)} 
^{d_i} e^{-\frac{1}{4\alpha't}\sum_i (y^i_1-y^i_2)^2}
\nonumber\\
&~&\times \frac{1}{q} \bigg{(}\prod_{n=1}^{\infty} \bigg{[} \bigg{(}
\frac{1-q^{2n}}{1+q^{2n-1}} \bigg{)}^2 \;\frac{\det(1+S_1S^T_2q^{2n-1})}
{\det(1-S_1S^T_2q^{2n})} \bigg{]}
\nonumber\\
&~&-\prod_{n=1}^{\infty} \bigg{[} \bigg{(} \frac{1-q^{2n}}{1-q^{2n-1}}
\bigg{)}^2 \;\frac{\det(1-S_1S^T_2q^{2n-1})}{\det(1-S_1S^T_2q^{2n})}
\bigg{]} \bigg{)} \;\bigg{\}} \;\;,
\eea
where $V_u$ is the common world volume of the two mixed branes.
\subsection{The R-R sector amplitude}
In the R-R sector there are 
\bea
_R \langle B^1_{\psi} , \eta_1 \mid \;= \langle A \mid \langle 
{\tilde B} \mid {\cal{N}}^{(\eta_1)}_{AB} e^{-i\eta_1 \sum_{m=1}
^{\infty}d^{\mu}_mS^{(1)}_{\mu \nu}{\tilde d}^{\nu}_m}\;\; ,
\eea
where ${\cal{N}}^{(\eta_1)}$ is given by 
\bea
{\cal{N}}^{(\eta_1)} = (-1)^{p_1}C\Gamma^0 \Gamma^{{\bar \alpha}_1}...
\Gamma^{{\bar \alpha}_{p_1}}G_1 \bigg{(} \frac{1-i\eta_1\Gamma_{11}}
{1+i\eta_1} \bigg{)} \;\;,
\eea
and
\bea
_R \langle B^1_{sgh} , \eta_1 \mid \;= \langle P=-3/2 , {\tilde P}
= -1/2 \mid e^{i\eta_1\beta_0{\tilde \gamma}_0-i\eta_1
\sum_{m=1}^{\infty}(\gamma_m {\tilde \beta}_m -\beta_m 
{\tilde \gamma}_m)}\;\;.
\eea

In calculation of ${\cal{A}}_{R-R}(\eta_1 , \eta_2) =\;_R \langle 
B_1 , \eta_1 \mid D \mid B_2 , \eta_2 \rangle _R $ we see that zero mode 
contribution of the superghost is 
\bea
_R ^{(0)} \langle B^1_{sgh} , \eta_1 \mid B^2_{sgh} , \eta_2 \rangle 
^{(0)}_R = \sum_{m=0}^\infty (\eta_1 \eta_2)^m \;\;.
\eea
 which for $\eta_1 \eta_2 =+1$ is divergent, and for $\eta_1 \eta_2 = -1$
is an alternating sum. This expression needs to be regularized to have 
a meaning. We introduce a special regularization scheme 
similar Ref.\cite{19}. For this we define
\bea
^{(0)} _R \langle B_1 , \eta_1 \mid B_2 , \eta_2 \rangle ^{(0)} _R
\equiv  \lim_{x \rightarrow 1}       
{}^{(0)} _R \langle B^1 _{sgh} , \eta_1 \mid                         
x^{2G_0} \mid B^2 _{sgh} , \eta_2 \rangle ^{(0)} _R 
\;{}^{(0)} _R \langle B^1 _{\psi} , \eta_1 \mid B^2 _{\psi} , \eta_2 
\rangle ^{(0)} _R \;\;,
\eea                                                              
similar to the equation (3.8) of Ref. \cite{7}. Also 
$G_0$ is defined in (50), i.e. $G_0= -\gamma_0 \beta_0$, therefore
\bea
^{(0)} _R \langle B^1_{sgh} , \eta_1 \mid x^{2G_0} \mid B^2_{sgh} , 
\eta_2 \rangle ^{(0)} _R = \frac{1}{1-\eta_1 \eta_2 x^2} \;\;.
\eea
For alternating sum (i.e. $\eta_1 \eta_2=-1$) this becomes $\frac{1}{1+x^2}$
, and for $x=1$ reduces to $\frac{1}{2}$. For $\eta_1 \eta_2=+1$ (i.e.
$\eta_1=\eta_2 \equiv \eta)$ is  
\bea
^{(0)} _R \langle B^1_{sgh} , \eta \mid x^{2G_0} \mid B^2_{sgh} , 
\eta \rangle ^{(0)} _R = \frac{1}{1- x^2} \;\;.
\eea
By an appropriate insertion of $\beta_0 , \gamma_0 , {\tilde \beta_0}$ and
${\tilde \gamma_0}$ in the left hand side of (63), projecting it out,
therefore
\bea
\lim_{x \rightarrow 1} {}^{(0)} _R \langle B^1 _{sgh} , \eta 
\mid x^{-2 \gamma_0 \beta_0} 
\delta (\beta_0-\frac{1}{4 \pi} \gamma_0) \delta ({\tilde \beta_0} + 
\frac{1}{4 \pi} {\tilde \gamma_0}) \mid B^2 _{sgh} , 
\eta \rangle ^{(0)} _R = 1 \;\;.
\eea
This gives a modified partition function for $\eta_1 = \eta_2$, which is
regular. Also zero mode part of the fermions becomes
\bea
{\cal{A}}^{0(R)}_{\psi}(\eta_1 , \eta_2 ) &\equiv&  
^{(0)} _R \langle B^1_{\psi} , \eta_1 \mid B^2_{\psi} , \eta_2 
\rangle ^{(0)}_R = {\rm Tr} \bigg{(} {\cal{M}}^{(\eta_2)}
C^{-1} {\cal{N}}^{(\eta_1)^T} C^{-1} \bigg{)} \;\;,
\eea
 this gives  
\bea
{\cal{A}}^{0(R)}_{\psi}(+ , - ) = {\cal{A}}^{0(R)}_{\psi}
(- ,+ ) =2 \zeta \;\;,
\eea
\bea
{\cal{A}}^{0(R)}_{\psi}(+ , + ) = {\cal{A}}^{0(R)}_{\psi}
(- ,- ) = \zeta' \;\;,
\eea
where $\zeta$ and $\zeta'$ have definition as
\bea
\zeta \equiv -\frac{1}{2} {\rm Tr} 
\bigg{[} G_1C^{-1}G^T_2C(\Gamma^{{\bar \beta}
_{p_2}} ... \Gamma^{{\bar \beta}_1})(\Gamma^{{\bar \alpha}_1} ... 
\Gamma^{{\bar \alpha}_{p_1}}) \bigg{]} \;\;, 
\eea
\bea
\zeta' \equiv i {\rm Tr} \bigg{[} G_1C^{-1}G^T_2C(\Gamma^{{\bar \beta}
_{p_2}} ... \Gamma^{{\bar \beta}_1})(\Gamma^{{\bar \alpha}_1} ... 
\Gamma^{{\bar \alpha}_{p_1}}) \Gamma_{11} \bigg{]} \;\;. 
\eea
With a simple algebra we see that $\zeta$ is symmetric and $\zeta'$
is antisymmetric under the 
exchange of the indices 1 and 2. To see this we use the $(\Gamma ^\mu)^T=
- C \Gamma ^\mu C^{-1} \;,\; C^T=-C $. 
Also we have $(\zeta'_{12})^*=\zeta'_{21}$ as needs for the  
symmetry of amplitude. 
For ${\cal{F}}_1={\cal{F}}_2=0$, we have $\zeta'=0$, therefore, with this
special regularization scheme
$\zeta'$ is purely the effect of the gauge fields. Adding all these 
together we obtain the contribution of zero modes,
\bea
{\cal{A}}^0 _{R-R} (\eta_1 , \eta_2) = \frac{1}{2} \bigg{(} (1-\eta_1
\eta_2)\zeta + (1+\eta_1 \eta_2)\zeta' \bigg{)} \;\;.
\eea
Note that we can write
\bea
C^{-1}G^T_2 C = e^{-\frac{1}{2}{\cal{F}}_{(2) \alpha \beta}\Gamma^ \alpha
\Gamma ^{\beta} }\;\;,
\eea
with the previous convention for the right hand side. It is worth emphasizing
that the right hand side is not $G^{-1}_2$. 
Therefore the R-R sector amplitude becomes
\bea
{\cal{A}}_{R-R} &=& \frac{T_{p_1}T_{p_2}}{8(2\pi)^{d_i}}
 \alpha' \int _0 ^{\infty} dt \bigg{\{} 
 \bigg{(} \sqrt{\frac{\pi}{\alpha' t}}\;\bigg{)} ^{d_{i_n}}
 e^{-\frac{1}{4\alpha' t}\sum_{i_n} (y^{i_n}_1 - y^{i_n}_2)^2} 
\nonumber\\
&~&\times \prod_{i_c} \Theta_3 \bigg{(} \frac{y^{i_c}_1 - y^{i_c}_2}
{2\pi R_{i_c}} \mid \frac{i\alpha' t}{\pi (R_{i_c})^2} \bigg{)}
\bigg{[} \bigg{[} \zeta \prod_{n=1}^{\infty} \bigg{[} \bigg{(} 
\frac{1-q^{2n}}{1+q^{2n}}
\bigg{)}^2 \; \frac{\det(1+S_1S^T_2q^{2n})}
{\det(1-S_1S^T_2q^{2n})} \bigg{]} + \zeta' \bigg{]} \bigg{]}
\nonumber\\
&~& \times \sum_{\{N^{u_c}\}} \bigg{[}(2\pi)^{d_u} 
\prod_u [\delta(p^u_1 -p^u_2)]  
\; \exp [\frac{i}{\alpha'}{\ell}^{u_c} ({\cal{F}}^{\alpha'_1}
_{(1)\;\;u_c}y^{\alpha'_1}_2 - {\cal{F}}^{\alpha'_2}_{(2)\;\;
u_c}y^{\alpha'_2}_1 )] 
\nonumber\\
&~&\times \exp [-\frac{t}{\alpha'} {\ell}^{u_c} {\ell}^{v_c}
(\eta_{u_c v_c}+{\cal{F}}^u_{(1)\;u_c}{\cal{F}}_{(2)\;uv_c}
+{\cal{F}}^{\alpha'_1}_{(1)\;u_c}{\cal{F}}^{\alpha'_1}_{(1)\;v_c}
+{\cal{F}}^{\alpha'_2}_{(2)\;u_c} {\cal{F}}^{\alpha'_2}_{(2)\;v_c}
)] \bigg{]}\;\bigg{\}} \;.  
\eea
We see that the signs of $\zeta$ and $\zeta'$ depend on the 
arrangements of ${\bar \alpha}_i$'s (and ${\bar 
\beta}_i$'s) in their arguments, therefore the R-R forces may be repulsive or attractive
due to the brane-brane or brane-antibrane interaction. Because of our 
special regularization, the 
quantity $\zeta'$ is usually zero. Due to their procedure, authors of 
Ref.\cite{7}, have non-zero $\zeta'$ for $D_0-D_8$ system. 
Some special configurations have non zero $\zeta'$,
for example consider $m_2$ and $m_8$-branes along $(X^1 , X^9)$ and
$(X^1 , ... , X^8)$ directions respectively, then
\bea
\zeta' = -i {\rm Tr} 
\bigg{(} G_1 C^{-1} G^T_2 C (\Gamma^9 ... \Gamma^2) \Gamma_
{11} \bigg{)} \;\;,
\eea
\bea
G_1 = 1+{\cal{F}}_{(1)01} \Gamma^0 \Gamma^1 + {\cal{F}}_{(1)09}
\Gamma^0 \Gamma^9 +{\cal{F}}_{(1)19} \Gamma^1 \Gamma^9 \;\;,
\eea
\bea
C^{-1}G^T_2 C = 1-{\cal{F}}_{(2)01} \Gamma^0 \Gamma^1 + ... \;\;,
\eea
therefore
\bea
\zeta' = 32i({\cal{F}}_{(2)01}-{\cal{F}}_{(1)01}) \;\;.
\eea
This result also hold for $m_3$ and $m_7$-branes along $(X^1, X^8,X^9)$
and $(X^1, ... ,X^7)$ directions.

Comparison of (55) and (72) says that the effects of compactification 
in ${\cal{A}}_{NS-NS}$ and in ${\cal{A}}_{R-R}$ are the same. 
This is due to the fact that upon compactification only bosonic 
contribution is modified. When these results are used in case of  
parallel mixed branes with the same dimension, those terms which contain
$\alpha'_1$ and $\alpha'_2$ disappear.
Again return to the non compact spacetime, therefore 
\bea
{\cal{A}}^{(nc)}_{R-R} &=& \frac{T_{p_1}T_{p_2}}{8(2\pi)^{d_i}}
 \alpha' V_u \int_0^{\infty} dt \bigg{\{}  \bigg{(}
\sqrt{\frac{\pi}{\alpha' t}}\;\bigg{)}^{d_i} e^{-\frac{1}{4\alpha't}
\sum_i (y^i_1 - y^i_2)^2}
\nonumber\\
&~&\times \bigg{[} \zeta \prod_{n=1}^{\infty} \bigg{[} \bigg{(} 
\frac{1-q^{2n}} 
{1+q^{2n}} \bigg{)}^2 \;\frac{\det(1+S_1S^T_2q^{2n})}{\det(1-S_1S^T_2q^{2n})}
\bigg{]} + \zeta' \bigg{]} \;\;\bigg{\}} \;\;.  
\eea                                                                        
\subsection{A special case}
Now we consider the important example of parallel branes with the
same ${\cal{F}}$.
Consider two parallel $m_{p_1}$ and $m_{p_2}$-branes which their world-branes
are at $(X^0,X^1,...,X^{p_1})$ and $(X^0,X^1,...,X^{p_1},...,X^{p_2})$ 
respectively with $\gamma \equiv p_2-p_1 \geq 0$. Also consider ${\cal{F}}
_{(1)\;uv}={\cal{F}}_{(2)\;uv}\equiv {\cal{F}}_{uv}$ for $u,v \in \{0,1,...,
p_1 \}$ and all other components of ${\cal{F}}_2$ be zero, therefore
orthogonality of $Q_{(1)\;uv} (=Q_{(2)\;uv})$ gives
\bea
\det(1+S_1S^T_2q_n)=(1+q_n)^{10-\gamma}(1-q_n)^{\gamma} \;\;,
\eea
where $q_n=\pm q^{2n} , \pm q^{2n-1}$. Also equality of the field strengths
implies $G_1=G_2 \equiv G$, therefore
\bea
\zeta= -\frac{1}{2} \delta_{\gamma ,0}{\rm Tr}
(GC^{-1}G^TC)=-16\delta_{\gamma ,0}
\det(1-{\cal{F}})\;\;,
\eea
the last equality can be investigated for each $``p_1$'' 
individually. For this special configuration equality of the field 
strengths implies $\zeta' = 0$. Finally
the total amplitude ${\cal{A}}={\cal{A}}_{NS-NS}+{\cal{A}}_{R-R}$ becomes
\bea
{\cal{A}} &=& \frac{T_{p_1}T_{p_2}\alpha'V_{p_1+1}}{8(2\pi)^{9-p_2}}
\det(1-{\cal{F}})\int_0^\infty dt \bigg{\{} \bigg{(} 
\sqrt{\frac{\pi}{\alpha't}} \;\bigg{)}^{d_{i_n}} 
e^{-\frac{1}{4\alpha't}\sum_{i_n}(y^{i_n}_1
-y^{i_n}_2)^2} 
\nonumber\\
&~&\times \prod_{i_c}\Theta_3 \bigg{(} \frac{y^{i_c}_1-y^{i_c}_2}
{2\pi R_{i_c}} \mid \frac{i\alpha't}{\pi (R_{i_c})^2}
\bigg{)} \bigg{(}\bigg{(}\; \frac{1}{q} \bigg{[}\bigg{[} 
\prod_{n=1}^{\infty} \bigg{[} \bigg{(}
\frac{1+q^{2n-1}}{1-q^{2n}} \bigg{)}^{8-\gamma}\bigg{(}\;\frac{1-q^{2n-1}}
{1+q^{2n}}\bigg{)}^{\gamma}\; \bigg{]}
\nonumber\\
&~&-\prod_{n=1}^{\infty} \bigg{[} \bigg{(} \frac{1-q^{2n-1}}{1-q^{2n}}
\bigg{)}^{8-\gamma} \bigg{(}\;\frac{1+q^{2n-1}}{1+q^{2n}}\bigg{)}^{\gamma}
\;\bigg{]}\;\;\; \bigg{]}\bigg{]} -16\delta_{\gamma ,0}\prod_{n=1}^{\infty}
\bigg{(} \frac{1+q^{2n}}{1-q^{2n}}\bigg{)}^{8-2\gamma} \bigg{)}\bigg{)}
\nonumber\\
&~&\times \sum_{\{N^{u_c}\}}\exp[-\frac{t}{\alpha'}{\ell}^{u_c}{\ell}^{v_c}
( \eta_{u_c v_c}+{\cal{F}}^u _{\;\;u_c} {\cal{F}}_{\;uv_c})]\; \bigg{\}}\;\;.  
\eea
Therefore the tensions are modified by the factor $\sqrt{\det(1-{\cal{F}})}$.
Apart from the modification of the tensions, 
field strengths ${\cal{F}}$ appear in this interaction amplitude 
only through the compactification effects.
 The first two terms come from the NS-NS sector, for $\gamma=p_2-p_1=4$
the NS-NS sector amplitude vanishes.
The third term comes from the R-R sector, and show that the amplitude of the 
R-R sector for the branes of different dimensions $(\gamma \neq 0)$
vanishes. For $\gamma=0$, total amplitude ${\cal{A}}$ vanishes (due
to the ``abstruse identity'') so the BPS no force condition is satisfied.

In non-compact spacetime,
making a transformation $t \rightarrow \pi/2t$
and for $T_p = \sqrt{\pi}(4\pi^2 \alpha')^{(3-p)/2}$ the last zero
amplitude transforms to the known parallel $D_p$-branes amplitude 
\cite{20} with the expected extra factor,
\bea
{\cal{A}} &=& V_{p+1} \det(1-{\cal{F}}) \int_0 ^{\infty} 
\frac{dt}{t} \bigg{\{}
(8\pi ^2 \alpha' t)^{-(p+1)/2} e^{-tY^2/2\pi \alpha'}
\prod_{n=1}^{\infty} (1-q^{2n})^{-8}
\nonumber\\
&~&\times \frac{1}{2} \bigg{[} \frac{1}{q} \bigg{(} \prod_{n=1}^{\infty} 
(1+q^{2n-1})^8 - \prod_{n=1}^{\infty} (1-q^{2n-1})^8 \bigg{)}
-16 \prod_{n=1} ^{\infty} (1+q^{2n})^8 \bigg{]} \bigg{\}} \;\;,
\eea
where $q=e^{-\pi t}$ and $Y^i=y^i_1-y^i_2$ is the separation of the 
branes.
\subsection{Other examples} 
In this part we give the interaction amplitude of the following
special systems. In these systems back-ground fields 
and effects of the compactification appear more explicitly. These systems
are : parallel $m_1-m_{1'}$-branes along $X^1$, $m_1$-brane along $X^1$
perpendicular to $m_{1'}$ along $X^2$, $m_0$-brane 
in front of $m_2$-brane along $X^1X^2$, parallel $m_2-m_{2'}$-branes
along $X^1X^2$, $m_2$-brane along $X^1X^2$ perpendicular to $m_{2'}$-brane 
along $X^2X^3$, $m_1$-brane along $X^1$ parallel to $m_5$-brane along
$X^1...X^5$ directions.
For all these we give the following amplitude
\bea
{\cal{A}} &=& \frac{T_pT_{p'}\alpha'V_u}{8(2\pi)^{d_i}}
\int_0^\infty dt \bigg{\{} \bigg{(} 
\sqrt{\frac{\pi}{\alpha't}} \;\bigg{)}^{d_{i_n}} 
e^{-\frac{1}{4\alpha't}\sum_{i_n}(y^{i_n}_1
-y^{i_n}_2)^2} 
\prod_{i_c}\Theta_3 \bigg{(} \frac{y^{i_c}_1-y^{i_c}_2}
{2\pi R_{i_c}} \mid \frac{i\alpha't}{\pi (R_{i_c})^2}
\bigg{)} 
\nonumber\\
&~&\times \bigg{(}\bigg{(}\sqrt{ff'}\; \frac{1}{q} \bigg{[}\bigg{[} 
\prod_{n=1}^{\infty} \bigg{[} \bigg{(}
\frac{1+q^{2n-1}}{1-q^{2n}} \bigg{)}^N \frac{w({\cal{F}},{\cal{F}}'
,q^{2n-1})}{w({\cal{F}},{\cal{F}}',-q^{2n})} \bigg{]}
-\prod_{n=1}^{\infty} \bigg{[} \bigg{(} \frac{1-q^{2n-1}}{1-q^{2n}}
\bigg{)}^N \frac{w({\cal{F}},{\cal{F}}',-q^{2n-1})}{w({\cal{F}},
{\cal{F}}',-q^{2n})} 
\bigg{]}\;\; \bigg{]}\bigg{]} 
\nonumber\\
&~&-16z \prod_{n=1}^{\infty}
\bigg{[} \bigg{(}\frac{1+q^{2n}}{1-q^{2n}} \bigg{)}^N \frac{w({\cal{F}},
{\cal{F}}',q^{2n})}{w({\cal{F}},{\cal{F}}',-q^{2n})} \bigg{]}
\;\;\bigg{)}\bigg{)} \theta({\cal{F}},{\cal{F}}',t,R,y) \bigg{\}} \;\;,
\eea
the parameters $V_u,d_i,f,f',N$ and $z$ for the above systems are collected
in the following table.

\begin{table}[ht]
\vspace{0.3cm}
\begin{center}
\begin{tabular}{|c|c|c|c|c|c|c|c|c|c|c|}
\hline
$p$&$p'$&$ $&$V_u$&$d_i$&${\cal{F}}$&${\cal{F}}'$&$f$&$f'$&$N$&$z$ \\
\hline
$1$&$1$&$\|$&$(2\pi R_1)L$&$8$&${\cal{F}}_{01}=E$&${\cal{F}}'_{01}=E'$&$
1-E^2$&$1-E'^2$&$6$&$1-EE'$\\
\hline
$1$&$1$&$\bot$&$L$&$7$&${\cal{F}}_{01}=E$&${\cal{F}}'_{02}=E'$&$1-E^2$&$
1-E'^2$&$5$&$-EE'$\\
\hline
$ $&$ $&$ $&$ $&$ $&${\cal{F}}_{01}=E_1$&$ $&$1-E_1^2$&$ $&$ $&$ $\\
$2$&$0$&$-$&$L$&$7$&${\cal{F}}_{02}=E_2$&$0$&$-E_2^2+B^2$&$1$&$5$&$-B$\\
$ $&$ $&$ $&$ $&$ $&${\cal{F}}_{12}=B$&$ $&$ $&$ $&$ $&$ $\\
\hline
$ $&$ $&$ $&$(2\pi R_1) \times$&$ $&${\cal{F}}_{01}=E_1$&${\cal{F}}'_{01}=
E'_1$&$ 1-E^2_1$&$1-E'^2_1$&$ $&$1-E_1 E'_1$\\
$2$&$2$&$\|$&$ (2\pi R_2)L$&$7$&${\cal{F}}_{02}=E_2$&${\cal{F}}'_{02}
=E'_2$&$-E_2^2+B^2$&$-E'^2_2+B'^2$&$5$&$-E_2E'_2+BB'$\\
$ $&$ $&$ $&$ $&$ $&$ {\cal{F}}_{12}=B$&${\cal{F}}'_{12}=B'$&$ $&$ $
&$ $&$ $\\ 
\hline
$ $&$ $&$  $&$ $&$ $&${\cal{F}}_{01}=E_1$&${\cal{F}}'_{02}=E'_2$
&$1-E_1^2$&$1-E'^2_2$&$ $&$ $\\
$2$&$2$&$\bot$&$(2\pi R_2)L$&$6$&${\cal{F}}_{02}=E_2$&${\cal{F}}'_{03}
=E_3 '$&$-E^2 _2 +B^2 $&$-E_3 '^2 +B'^2 $&$4$&$E_1E'_3+BB'$\\
$ $&$ $&$ $&$ $&$ $&${\cal{F}}_{12}=B$&${\cal{F}}'_{23}=B'$&$ $&$ $&$ $
&$ $\\
\hline
$ $&$ $&$  $&$ $&$ $&${\cal{F}}_{01}=E_1$&$ $&$1-E^2_1$&$ $&$ $&$ $\\
$5$&$1$&$\|$&$(2\pi R_1)L$&$4$&${\cal{F}}_{02}=E_2$&${\cal{F}}'_{01}=E'_1$
&$-E^2_2+B^2$&$1-E'^2_1$&$2$&$0$\\
$ $&$ $&$ $&$ $&$ $&${\cal{F}}_{12}=B$&$ $&$ $&$ $&$ $&$ $\\
\hline
\end{tabular}
\end{center}
\end{table}

 Note that ``$\|$'' and ``$\bot$'' stand for the 
 ``parallel'' and ``perpendicular'' respectively,   
 and $L$ is infinite time length.

Now we give the functions $\theta({\cal{F}},{\cal{F}}',t,R,y)$ and
$w({\cal{F}},{\cal{F}}',q_n)$ for these systems, therefore more properties 
of the interaction of these systems will become clear.


{\bf Parallel ${\bf m_1}$-branes}

For this system we have
\bea
\theta(E,E',t,R_1)=\frac{2\pi}{L}\sum_{m=-\infty}^{\infty}\delta[(E-E')
mR_1/{\alpha'}]e^{-t(1-EE')m^2R_1^2/{\alpha'}} \;\;,
\eea
where $R_1$ is the radius of compactification of $X^1$, therefore
$V_2=(2\pi R_1)L$. Also $Q$ is given by the matrix
\bea
Q=\left( \begin{array}{cc}
\frac{1+E^2}{1-E^2} & -\frac{2E}{1-E^2} \\
-\frac{2E}{1-E^2} & \frac{1+E^2}{1-E^2}
\end{array} \right) \;\;,
\eea
and $Q'$ has the same form as $Q$ in which $E$ is replaced by $E'$. We also
have,
\bea
w(E,E',q_n)=\det(1+q_nQQ'^T) \;\;.
\eea
Note that to get $Q'^T$ from $Q'$, one must use of 
\bea
(Q'^T)^{\alpha}\;\;_{\beta}=(Q')_{\beta}\;\;^{\alpha}=\eta ^{\alpha \alpha}
\eta _{\beta \beta}(Q')^{\beta}\;\;_{\alpha}\;\;,
\eea
therefore
\bea
w(E,E',q_n)=\bigg{(} 1+\frac{(1-E)(1+E')}{(1+E)(1-E')}q_n \bigg{)}
\bigg{(}1+\frac{(1+E)(1-E')}{(1-E)(1+E')}q_n \bigg{)} \;\;.
\eea
For $E=E'$ we have
\bea
\theta(E,E,t,R_1)=\Theta_3\bigg{(}0 \mid \frac{it(1-E^2)R_1^2}{\pi \alpha'}
\bigg{)} \;\;,
\eea
therefore through the compactification, fields $E=E'$ appear in the 
amplitude as in (88) (except for the factors $\sqrt{1-E^2}$ in the 
modification of the
tensions). For $E=E'$ the amplitude vanishes (due to the abstruse identity).

{\bf Perpendicular ${\bf m_1}$-branes}

In this case
\bea
\theta(E,E',t,R_{\alpha})=1 \;\;,
\eea
also $w(E,E',q_n)=\det(1+q_n \Omega \Omega'^T)$ where $\Omega$ and $\Omega'$
are
\bea
\Omega = \left( \begin{array}{cc}
Q & 0 \\
0 & -1
\end{array} \right) \;\;,
\eea
\bea
\Omega'=\left( \begin{array}{ccc}
\frac{1+E'^2}{1-E'^2} & 0 & -\frac{2E'}{1-E'^2} \\
0 & -1 & 0 \\
-\frac{2E'}{1-E'^2} & 0 & \frac{1+E'^2}{1-E'^2}
\end{array} \right) \;\;,
\eea
where $Q$ is the same as in (84). After the expansion of the determinant we
see that the function $w(E,E',q_n)$ is symmetric under the exchange of $E$ 
and $E'$, as expected, 
For this system $z=EE'$, therefore R-R interaction may be attractive,
 repulsive or zero, according to the signs and values of $E$ and $E'$. We 
 remind that the function $w$ simplifies to,

\bea
w(E,0,q_n)=(1-q_n)^2 (1+q_n) \;\;.
\eea

{\bf ${\bf m_2-m_0}$ Branes system}

For this system the functions $\theta$ and $w$ are
\bea
\theta({\cal{F}},{\cal{F}}',t,R_{\alpha})=1 \;\;,
\eea
\bea
w(E_1,E_2,B,q_n)=\det(1+q_nQ{\Omega '}^T) \;\;,
\eea
where $Q$ and $\Omega'$ are $3 \times 3$ matrices 
\bea
\Omega '= {\rm diag}(1,\;-1,\;-1) \;\;,
\eea
\bea
Q=\frac{1}{f}\left( \begin{array}{ccc}
(1+E_1^2+E_2^2+B^2) & 2(-E_1+E_2 B) & 
-2(E_2 + E_1 B) \\ 
-2(E_1+E_2 B) & (1+E_1 ^2 -E_2 ^2-B^2)
& 2(B+E_1 E_2) \\ 
2(-E_2+E_1 B) & 2(-B+E_1 E_2) &
(1-E_1 ^2 + E_2 ^2-B^2)
\end{array} \right) \;\;,
\eea
and $f=1-E_1^2-E_2^2+B^2$. After the expansion of the 
determinant we see that $w(E_1,E_2,B,q_n)$
is symmetric under the exchange of the $E_1$ and $E_2$ as expected. Also
$z=-B$ says that the R-R interaction is attractive for positive 
${\cal{F}}_{12}=B$ and is repulsive for negative $B$, in other word
R-R force depends on the fact that $m_0$-brane is in what sides of
$m_2$-brane.

{\bf Parallel ${\bf m_2}$-branes}

In this case again we can write the function $w$ as
$w({\cal{F}},{\cal{F}}',q_n)=\det(1+q_n Q{Q'}^T)$,
where the matrix $Q$ is given in (96). The matrix $Q'$ has exactly 
the same form of the matrix $Q$ with $E_1,E_2$ and $B$ changed to
$E'_1,E'_2$ and $B'$ respectively. (We remind that $(Q'^T)^{\alpha}\;\;
_{\beta}={\eta}^{\alpha \alpha}{\eta}_{\beta \beta}(Q')^{\beta}\;\;
_{\alpha}$).

Now consider the case where $X^1$ and $X^2$-directions are 
both compact. Therefore
\bea
\theta({\cal{F}},{\cal{F}}',t,R_1,R_2) &=& \frac{(2\pi)^3}{V_3} \sum_{m=
-\infty}^{\infty}\sum_{n=-\infty}^{\infty} \bigg{\{} \delta \bigg{(}
(E_1-E'_1)mR_1/{\alpha'}+(E_2-E'_2)nR_2/{\alpha'} \bigg{)}
\nonumber\\
&~& \times \delta [(B-B')mR_1/{\alpha'}] \delta[(B-B')nR_2/{\alpha'}]
\exp \bigg{[} -\frac{t}{\alpha'} \bigg{(} (1-E_1 E'_1+BB')m^2R_1^2
\nonumber\\
&~& +(1-E_2E'_2+BB')n^2R_2^2-(E_1E'_2+E'_1E_2)mnR_1R_2 \bigg{)} \bigg{]}
\bigg{\}}\;\;,
\eea
where $V_3 = (2\pi R_1)(2\pi R_2)L$ is the world volume of the $m_2$-branes.
We see that the functions $w$ and $\theta$ are symmetric 
under the exchange of the fields ${\cal{F}}$ and ${\cal{F}}'$, as expected.
 Specially consider $E'_1=E_1 ,\; E'_2=E_2$ and $B'=B$, which give 
$QQ'^T=1$. Therefore except for
the factors $\sqrt{1-E_1^2-E_2^2+B^2}$ in the modification of the tensions,
compactification causes that these fields to appear in the amplitude by 
the equation (97).
More specially let $E_2=E'_2=0$ then
\bea
\theta(E_1,B,t,R_1,R_2)=\Theta_3 \bigg{(} 0 \mid \frac{it(1-E_1^2+B^2)R_1^2)}
{\pi \alpha'} \bigg{)} \Theta_3 \bigg{(} 0 \mid \frac{it(1+B^2)R_2^2}
{\pi \alpha'} \bigg{)} \;\;.
\eea

{\bf Perpendicular ${\bf m_2}$-branes}

Consider $m_2$-brane along the $(X^1,X^2)$ and $m_{2'}$-brane 
along $(X^2,X^3)$
directions, then $w({\cal{F}},{\cal{F}}',q_n)=\det(1+q_n \Omega 
\Omega'^T)$, where $\Omega$ and $\Omega'$ are
\bea
\Omega = \left( \begin{array}{cc}
Q & 0 \\
0 & -1 
\end{array} \right) \;\;,
\eea
\bea
\Omega' = \left( \begin{array}{cccc}
\frac{1}{f'}(1+E_2'^2 +E_3'^2 + B'^2 ) & 0 & \frac{2}{f'}(-E_2' +E_3' B') & 
-\frac{2}{f'}(E_3' +E_2' B') \\
0 & -1 & 0 & 0 \\
-\frac{2}{f'}(E_2'+E_3' B') & 0 & \frac{1}{f'}(1+E_2'^2 -E_3'^2 -B'^2 ) & 
\frac{2}{f'}(B'+E_2' E_3' ) \\
\frac{2}{f'}(-E_3' +E_2' B') & 0 & \frac{2}{f'}(-B'+E_2' E_3' ) & 
\frac{1}{f'}(1-E_2'^2 +E_3'^2 -B'^2 )
\end{array} \right) \;\;,
\eea
where $Q$ is given in (96) and $f'=1-E'^2_1-E'^2_2+B'^2 $. 
The function $\theta$ is
\bea
\theta(E_2,B,E'_2,B',y_1^3,y_2^1,t,R_2) = \frac{2\pi}{L} \sum_{m=-\infty}
^{\infty} \bigg{\{} \delta[(E_2-E'_2)mR_2/{\alpha'}]
\nonumber\\
\times \exp \bigg{(}\frac{i}{\alpha'}mR_2(By_2^1+B'y_1^3)-\frac{t}{\alpha'}
(1-E_2E'_2+B^2+B'^2)m^2 R_2^2 \bigg{)} \bigg{\}} \;\;,
\eea
for $E_2 \neq E'_2$ we have $\theta=1$, but for $E_2= E'_2$ it is
\bea
\theta(E_2,B,E_2,B',y_1^3,y_2^1,t,R_2) = \Theta_3 \bigg{(} \frac{(By_2^1+
B'y_1^3)R_2}{2\pi \alpha'} \mid \frac{it(1-E_2^2+B^2+B'^2)R_2^2}{\pi \alpha'}
\bigg{)} \;\;.
\eea

{\bf ${\bf m_5}$-brane parallel to ${\bf m_1}$-brane}

For simplicity consider ${\cal{F}}_{01}=E_1,\; {\cal{F}}_{02}=E_2,\;
{\cal{F}}_{12}=B$ and all other components of ${\cal{F}}_{\alpha \beta}$
be zero, these with ${\cal{F}}'_{01}=E'_1$ give
\bea
w({\cal{F}},{\cal{F}}',q_n)=(1-q_n)^3 \det(1+q_n \Omega \Omega'^T) \;\;,
\eea
where $\Omega$ is the same as $Q$ in (96) and $\Omega'$ is given by  
(90) in which $E$ must change to $E'_1$.
 The function $\theta$ is
\bea
\theta(E_1,B,E'_1,t,R_1)=\frac{2\pi}{L} \sum_{m=-\infty}^{\infty}
\delta[(E_1-E'_1)mR_1/{\alpha'}] 
\nonumber\\
\times \exp \bigg{(} -\frac{i}{\alpha'}mR_1By_2^2
-\frac{t}{\alpha'}m^2R_1^2(1-E_1E'_1+B^2) \bigg{)} \;\;.
\eea
Note that for $E_1\neq E'_1$ it is equal to 1, and for $E_1=E'_1$ is given
by Jacobi function,
\bea
\theta(E_1,B,E_1,t,R_1)=\Theta_3 \bigg{(} -\frac{R_1By_2^2}{2\pi \alpha'}
\mid \frac{it(1-E_1^2+B^2)R_1^2}{\pi \alpha'} \bigg{)} \;\;.
\eea
For $E'_1=E_1$ and $E_2=B=0$, NS-NS interaction vanishes.
 R-R interaction of this system for any ${\cal{F}}$ and ${\cal{F}}'$ is 
always zero.

\subsection{Massless states contribution to the amplitude}
For distant branes only massless states have a considerable contribution
on the interaction amplitude . As NS-NS sector and R-R sector
massless states have zero momentum numbers 
and winding numbers, in equations (55) and (72), only the term
with $N^{u_c} = 0$ ( for all $ u_c$ ) contributes to these states. In
addition we must calculate the following limit
\bea
\Omega _{NS} &\equiv& \lim_{q \rightarrow 0}\;\frac{1}{q} \bigg{\{}
\prod_{n=1}^{\infty} \bigg{[} \bigg{(} \frac{1-q^{2n}}{1+q^{2n-1}}
\bigg{)}^2\; \frac{\det(1+Sq^{2n-1})}{\det(1-Sq^{2n})} \bigg{]}
\nonumber\\
&~&- \prod_{n=1}^{\infty} \bigg{[} \bigg{(} \frac{1-q^{2n}}{1-q^{2n-1}}
\bigg{)}^2 \; \frac{\det(1-Sq^{2n-1})}{\det(1-Sq^{2n})} \bigg{]} 
\bigg{\}}\;\;,
\eea
for the NS-NS sector and
\bea
\Omega_R &\equiv& \lim_{q \rightarrow 0} \;\prod_{n=1}^{\infty}
\bigg{[} \bigg{(} \frac{1-q^{2n}}{1+q^{2n}} \bigg{)}^2\;
\frac{\det(1+Sq^{2n})}{\det(1-Sq^{2n})}\bigg{]} \;\;,
\eea
for the R-R sector, where $q=e^{-2t}$ and $S=S_1S^T_2$. 
For a matrix $A$ we have $detA=e^{Tr[lnA]}$ therefore
\bea
\prod_{n=1}^{\infty} \bigg{(} \det(1+q_nS') \bigg{)}\; =\;\exp \bigg{\{} 
\sum_{k=0}^{\infty} \bigg{[} \frac{(-1)^k Tr(S'^{k+1})}{k+1} 
\sum_{n=1}^{\infty} q^{k+1}_n \bigg{]} \bigg{\}} \;\;,
\eea
where $q_n = q^{2n}$ , $q^{2n-1}$ and $S' = \pm S, \pm 1$, thus,
\bea
{\cal{A}}^{(NS-NS)}_0 &=& \frac{T_{p_1}T_{p_2}}{4(2\pi)^{d_i}}{\alpha'}
V_u \sqrt{\det(1-{\cal{F}}_1)\det(1-{\cal{F}}_2)}\;[Tr(S_1S^T_2)-2]
\nonumber\\
&~&\times \int_0^\infty dt \bigg{\{} \bigg{(} \sqrt{\frac{\pi}{\alpha' t}}
\;\bigg{)}^{d_{i_n}}e^{-\frac{1}{4\alpha' t}\sum_{i_n}(y^{i_n}_1
-y^{i_n}_2)^2} \prod_{i_c} \Theta_3 \bigg{(} \frac{y^{i_c}_1-y^{i_c}_2}
{2\pi R_{i_c}} \mid \frac{i\alpha' t}{\pi (R_{i_c})^2} \bigg{)} \bigg{\}}\;\;. 
\eea
For the system that was considered in subsection 3.3, and for $\gamma=4$ this
vanishes, meaning that the attractive force of graviton and dilaton cancel
the repulsive force of Kalb-Ramond field. For this case
\bea
{\cal{A}}^{(R-R)}_0 = \frac{T_{p_1}T_{p_2}}{8(2\pi)^{d_i}}
(\zeta + \zeta')
\alpha' V_u \int_0^{\infty} dt \bigg{\{} \bigg{(} \sqrt {\frac{\pi}
{\alpha' t}}\;\bigg{)}^{d_{i_n}} 
\nonumber\\
\times e^{-\frac{1}{4\alpha't}\sum_{i_n}(y^{i_n}_1-y^{i_n}_2)^2}
\prod_{i_c} \Theta_3 \bigg{(} \frac{y^{i_c}_1-y^{i_c}_2}
{2\pi R_{i_c}} \mid \frac{i\alpha't}{\pi(R_{i_c})^2} \bigg{)} \bigg{\}}\;\;.
\eea
Again for the system of subsection 3.3, for $\gamma \neq 0$ this always
is zero. 

For parallel $m_p$-branes ( or anti $m_p$-branes ) with ${\cal{F}}_1
={\cal{F}}_2 \equiv {\cal{F}} $ in non compact
space time the total massless states amplitude $\bigg{(} {\cal{A}}_0
={\cal{A}}^{(NS-NS)}_0 + {\cal{A}}^{(R-R)}_0 \bigg{)}$ is
\bea
{\cal{A}}_0 = (1-1)V_{p+1} 2T^2_p G_{9-p}(Y^2)\det(1-{\cal{F}}) \;\;,
\eea
where $Y^i = y^i_1-y^i_2$ and $G_D(Y^2)$ is the Green's function in $D$
dimension. For $T_p = \sqrt{\pi} (4\pi^2 \alpha')^{(3-p)/2}$, quantity
${\cal{A}}_0$ agrees with the known cases in the literatures with the
expected extra factor $\det(1-{\cal{F}})$.

\section{Conclusion}
We explicitly showed that how the total field strength ${\cal{F}}$ and 
compactification effects appear in the boundary states. A novel feature 
is to cause closed string states to have a momentum along the brane, where
the branes are wrapped on the compact directions.

We obtained the general form of the amplitude for branes with arbitrary 
dimensions $p_1 , p_2$ and internal field strengths ${\cal{F}}_1$ and
${\cal{F}}_2$ for both compact and non-compact spaces.
The sign of the zero mode part of R-R sector amplitude 
$\zeta$ and $\zeta'$ corresponds to the brane-brane (antibrane-antibrane)
or brane-antibrane interactions. For parallel mixed branes with the same
total field strength, only in the compactified space the field
strength appears in the interaction amplitude (except for the factors 
$\sqrt{\det(1-{\cal{F}})}$ in the modification of the tensions).
For this system when $p_2-p_1=4$, the NS-NS interaction vanishes, for 
$p_1 =p_2$ total interaction amplitude is zero, so the BPS no force 
condition is satisfied.

{\bf Acknowledgement}

The authors would like to thank M.M. Sheikh-Jabbari for useful discussion. 


\end{document}